\documentclass[twocolumn]{autart}

\usepackage{amsmath,amsfonts,amssymb,cases}
\usepackage{graphicx}

\newcommand{\be}{\begin{equation}}
\newcommand{\ee}{\end{equation}}
\newcommand{\bd}{\begin{displaymath}}
\newcommand{\ed}{\end{displaymath}}
\newcommand{\bea}{\begin{eqnarray}}
\newcommand{\eea}{\end{eqnarray}}

\newcommand{\R}{\mathbb{R}}
\newcommand{\C}{\mathbb{C}}
\newcommand{\Z}{\mathbb{Z}}

\newcommand{\Cs}{\mathbb{C}_{\sigma_0}}

\newcommand{\w}{\omega}
\newcommand{\so}{\sigma_0}
\newcommand{\dwz}{\Delta\w_{zr}}
\newcommand{\dwp}{\Delta\w_{pi}}
\newcommand{\dsz}{\Delta \sigma_{zr}}
\newcommand{\dsp}{\Delta \sigma_{pi}}
\newcommand{\N}{\mathbb{N}}

\begin{document}
\begin{frontmatter}

\title{Root Locus for SISO Dead-Time Systems: \\ A Continuation Based Approach}

\author[First]{Suat Gumussoy},
\author[First]{Wim Michiels}

\address[First]{Department of Computer Science, K. U. Leuven, \\
        Celestijnenlaan 200A, 3001, Heverlee, Belgium \\
        \mbox{(e-mails: \{suat.gumussoy, wim.michiels\}@cs.kuleuven.be)}.}

\begin{abstract}
We present a numerical method to plot the root locus of Single-Input-Single-Output (SISO) dead-time systems with respect to the controller gain or the system delay. We compute the trajectories of characteristic roots of the closed-loop system on a prescribed complex right half-plane. We calculate the starting, branch and boundary crossing roots of root-locus branches inside the region. We compute the root locus of each characteristic root based on a predictor-corrector type continuation method. To avoid the high sensitivity of roots with respect to the locus parameter in the neighborhood of branch points, the continuation method relies on a natural parameterization of the root-locus trajectory in terms of a distance in the (characteristic root, locus parameter)-space. The method is numerically stable for high order SISO dead-time systems.
\end{abstract}

\begin{keyword}
Root locus, dead time, time delay, SISO, characteristic root, stability analysis, relative stability.
\end{keyword}

\end{frontmatter}

\section{Introduction}
The root locus method is an essential tool in modern control engineering for analysis and synthesis problems \cite{OgataBook}. This method is successfully implemented for finite dimensional SISO systems and becomes a fundamental tool in control education \cite{Evans04,Krajewski07}.

The closed-loop of the SISO system with a time delay generically has infinitely many poles in the complex plane (see, e.g., \cite{WimBook}). Therefore obtaining the root locus for dead-time systems is a difficult problem. Unlike the finite dimensional case, the root-locus equation contains a time delay term and standard polynomial root-finding algorithms for the root computation are not applicable.

The root locus of SISO dead-time systems with respect to the \emph{controller gain} is based on two main approaches: methods based on sweeping or gridding parameters on the complex plane and continuation based methods.

The methods in the first group are early methods in the literature. The root locus is obtained on a set of  vertical lines \cite{HuangTAC67}, horizontal lines \cite{Yeung82} or on a rectangular grid \cite{Krall70} in the complex plane by finding points satisfying the root-locus equation. These methods require a large number of grid points for an accurate root-locus plot and may miss the dynamics of some characteristic roots due to the finite number of grid or sweep parameters.

The methods in the second group are continuation based methods. These methods follow characteristic roots of the root locus by a predictor-corrector algorithm inside the root-locus region and detect the ones entering into the region due to the asymptotic root chains of time delay systems. The predicted characteristic root is obtained by computing the slope of the phase equation of the root-locus equation \cite{AshTAC68}, the solution of nonlinear differential equations \cite{Suh82} or by a triangulation method on the complex plane \cite{Nishioka91} and the predicted values are corrected by a Newton-Raphson iteration. The characteristic roots entering into the root-locus region are detected by checking the sign of function values on the constant grid points of the region's boundary \cite{AshTAC68} or computing \emph{another} root locus problem \cite{Nishioka91,Suh82}. The approach proposed in \cite{AshTAC68} has two main limitations. First, the root locus is computed on a rectangular region and this is restrictive to observe the overall dynamics of characteristic roots. Second, it requires exhaustive search on a grid for characteristic roots entering into the root-locus region, with a possibility of missing a characteristic root depending on the number of grid points. The articles \cite{Nishioka91,Suh82} consider general SISO time delay systems, i.e., SISO systems with state delays. The predictor step in both methods, the solution of nonlinear differential equations and the triangulation method, are numerically expensive compared to the simple linear predictor. In order to detect asymptotic roots another root locus problem has to be solved and the number of required roots for a complex region is not given.

The root locus of SISO dead-time systems with respect to the \emph{time delay} allows us to analyze the effects of time delay on system stability and performances and it is less considered in the literature. A computationally expensive brute force approach is to compute the locations of the characteristic roots inside a desired region for a grid over the parameter space using, e.g., a spectral method \cite{BredaSISC2005}. A continuation based method in \cite{FioravantiTDS2010} obtains the root locus on the \emph{complex right half-plane} by solving a nonlinear differential equation and correcting with Newton's method. The predictor step is computationally expensive and the method does not allow to analyze the root locus of stable closed-loop systems since the boundary of the root-locus region is the imaginary axis.

All continuation based methods for the root locus of SISO dead-time systems in the existing literature parameterize root-locus trajectories with respect to the controller gain or the time delay. This parameterization is numerically ill-posed due to the high sensitivity in the neighborhood of intersection points characterized by the presence of multiple roots, as reported in \cite{FioravantiTDS2010}. Both the predictor and corrector steps in all methods require evaluations of functions with time delays. These evaluations are numerically expensive and not numerically stable due to the oscillation and exponential increase of time delay terms in the imaginary axis and the positive real axis directions in the complex plane respectively.

In this article, we compute the root locus of SISO dead-time systems on a \emph{given} complex right half-plane up to a predefined controller gain or a time delay. Our main contributions are the following:
\begin{itemize}
\item We compute the trajectories of \emph{all characteristic roots} entering into the prescribed region with respect to the controller gain or the time delay up to an upper bound value. By choosing the controller gain or time delay sufficiently large, the asymptotic behavior of the roots can be observed.
\item Our continuation method estimates the next root on the complex plane by a simple linear predictor and corrects this prediction with Newton's method. We parameterize the root-locus trajectories in the (root and gain/delay)-space in terms of the arclength which is numerically robust, as we shall see.
\item We use an \emph{adaptive step size} in the prediction step depending on the convergence rate of Newton's method and the distance of the root from the root-locus trajectory. This makes our algorithm scalable since it uses different step sizes for different root-locus trajectories and adapts to the curvature of the trajectories.
\item We transform the root-locus equation into phase and logarithmic magnitude equations and avoid the oscillation and exponential increase problems due to the time delay term in function evaluations, which are only needed in the correction step.
\end{itemize}

The paper is organized as follows. Section \ref{sec:prob} formulates the root-locus problem with respect to the controller gain and the time delay. The critical points of the root locus are computed in Section \ref{sec:critpts}. The predictor-corrector based continuation method is given in Section \ref{sec:rloci}. The overall algorithm for the root locus is presented in Section \ref{sec:alg}. Section~\ref{sec:ex} is devoted to numerical examples. In Section~\ref{sec:concl}, some concluding remarks are presented.

\textbf{Notation:}
\begin{tabbing}
  \= $\C, \R, \Z$ \=: sets of complex, real and integer numbers, \\
  \> $\Re(u)$ \>: real part of a complex number $u$, \\
  \> $\Im(u)$ \>: imaginary part of a complex number $u$, \\
  \> $|u|, \angle u$ \>: magnitude, phase of a complex number $u$, \\
  \> $\lceil u \rceil$ \>: the smallest integer larger than or equal to $u$,  \\
  \> $\lfloor u \rfloor$ \>: the largest integer smaller than or equal to $u$, \\
  \> $u^T$ \>: the transpose of the vector $u$, \\
  \> $A^+$ \>: the pseudoinverse of the matrix $A$, \\
  \> $\textrm{sgn}(u)$ \>: returns $+1, -1, 0$ given a real number $u$ \\
  \>  \>$\ $ for $u>0, u<0, u=0$ respectively, \\
  \> $\N_{n_1}^{n_2}$ \>: the set of integers from $n_1$ to $n_2$.   \\
\end{tabbing} \vspace{-0.72cm}
$\textrm{mod}(u,v)$: returns the remainder on division of the dividend $u$ by the divisor $v$.

\section{Problem Formulation} \label{sec:prob}
A SISO dead-time system is a rational, proper SISO plant $G$ with a constant input or output time delay $h$. The only required input data to compute the root locus of these systems with respect to the controller gain or the time delay are
\begin{itemize}
\item the poles of $G$, $p_i\in\C$ for $i\in\N_1^n$,
\item the zeros of $G$, $z_r\in\C$ for $r\in\N_1^m$,
\item the gain of $G$, $\alpha\in\R$,
\item the time delay $h$, $h\in\R$ and $h>0$.
\end{itemize}
Note that the input data can be obtained from state-space matrices or the transfer function of $G$.

Let $G(s)e^{-hs}$ be the transfer function representation of the SISO dead-time system where
\be \label{eq:tfG}
G(s)=\alpha\frac{(s-z_1)(s-z_2)\ldots(s-z_m)}
{(s-p_1)(s-p_2)\ldots(s-p_n)},\ \ n\ge m
\ee and the real and imaginary parts of system zeros and poles are defined as $z_r=\sigma_{zr}+j\omega_{zr}$ for $r\in\N_1^m$ and $p_i=\sigma_{pi}+j\omega_{pi}$ for $i\in\N_1^n$.

The \emph{root-locus equation} of a SISO dead-time system with respect to the controller gain or the time delay is
\begin{numcases}{f(s,\lambda)=0\ \ \textrm{where}\ \ f(s,\lambda)=}
   \label{eq:rlocusK} 1+\lambda G(s)e^{-hs} \\
   \label{eq:rlocusH} 1+G(s)e^{-\lambda s}
\end{numcases}
and $\lambda$ is the \emph{locus parameter} for $\lambda\in[0,\lambda_{\max}]$ where $\lambda_{\max}$ is a given positive number. The \emph{root-locus region} is a complex right half-plane
\be \label{eq:Cs}
\Cs =\left\{s\in \mathbb{C}: \Re(s)\geq\sigma_0\right\}
\ee where $\sigma_0$ is a negative real number. The corresponding \emph{boundary} of the root-locus region is a vertical line through $s=\so$ and parallel to the imaginary axis.

We define root-locus problems with respect to the controller gain or the time delay as follows.

\noindent \textbf{Root locus problems:} Given $\lambda_{\max}>0$, compute the locations of characteristic roots of root-locus equations in (\ref{eq:rlocusK},\ref{eq:rlocusH}) inside the root-locus region $\Cs$ for $\lambda\in[0,\lambda_{\max}]$.

 The \emph{root-locus trajectory} of the characteristic root is a curve on the complex plane on which each point satisfies the root-locus equation for $\lambda\in[0,\lambda_{\max}]$.The intersection of two or more root-locus trajectories is a \emph{branch point}. The \emph{starting points} of the root locus are the characteristic roots inside $\Cs$ for $\lambda=0$ and the \emph{boundary crossing roots} are the characteristic roots crossing $\Re(s)=\so$ for any non-negative $\lambda$ value smaller then $\lambda_{\max}$. The \emph{critical points} are the starting points, branch points and boundary crossing roots of the root locus.

For the root locus w.r.t.~the controller gain, we can compute branch points inside $\Cs$ apriori as we shall see. For the root locus w.r.t. the time delay,  we detect the branch points  while following the root-locus trajectory. The behavior of a root-locus trajectory around a branch point is given in the following Lemma \cite{Suh82}.
\begin{lem} \label{lem:branch}
If $\tilde{s}$ is a root of the root-locus equation in (\ref{eq:rlocusK}) or (\ref{eq:rlocusH}) with a locus parameter $\tilde{\lambda}$ satisfying
\bd
{\textstyle
\left.\frac{\partial^l f(s,\tilde{\lambda})}{\partial s}\right|_{s=\tilde{s}}=0, \ l\in\N_1^{N-1}\ \textrm{and} \ \left.\frac{\partial^N f(s,\tilde{\lambda})}{\partial s}\right|_{s=\tilde{s}}\neq0},
\ed
then $N$ root-locus trajectories intersect at $s=\tilde{s}$. Intersecting trajectories continue straight after the intersection if $N$ is odd and make an angle of $-\frac{\pi}{N}$ angle if $N$ is even.
\end{lem}

When a single root $s$ of the root-locus equation crosses the boundary \mbox{$\Re(s)=\so$}, we determine whether it enters into or leaves the region $\Cs$ by computing its \emph{crossing direction} \cite{WimBook} defined as
\be \label{eq:RT}
{\textstyle
\mathcal{CD}(s,\lambda):=\textrm{sgn}\left(\Re{\left(\left.-\frac{\frac{\partial f}{\partial \lambda}}{\frac{\partial f}{\partial s}}\right|_{f(s,\lambda)=0}\right)}\right)}.
\ee A root on the boundary enters into or leaves the region $\Cs$ depending on whether $\mathcal{CD}(s,\lambda)>0$ or $\mathcal{CD}(s,\lambda)<0$ respectively.

The computation of root-locus trajectories involves two main tasks: computing the critical points and following the root-locus trajectories. The next section focuses on the computation of the critical points of the root locus w.r.t. the controller gain and the time delay. We follow each root trajectory using a continuation method described in Section \ref{sec:rloci}.

\section{Computation of Critical Points} \label{sec:critpts}
We compute the critical points of two root-locus problems in the next two subsections.

\subsection{Controller gain as locus parameter} \label{sec:critptsK}
The root-locus equation w.r.t. the controller gain is given in (\ref{eq:rlocusK}). The starting points of the root locus are the poles of $G$ inside the root-locus region $\Cs$.

The branch points satisfy the root-locus equation and
\be
\frac{\partial f(s,\lambda)}{\partial s}=\lambda(G'(s)-G(s)h)e^{-hs}=0.
\ee
Thus the branch points are the zeros of the transfer function $G'(s)-G(s)h$ inside the region $\Cs$. This transfer function can be written as
\[
\frac{G'(s)}{G(s)}-h=\sum_{r=1}^m\frac{1}{s-z_r}-\sum_{i=1}^m\frac{1}{s-p_i}-h
\] and realized in a state-space representation as a series connection of $m+n$ first-order systems using the poles and zeros of $G$ and the time delay $h$. Its zeros can be computed by solving an eigenvalue problem constructed from state-space matrices of $G$. Consequently, the branch points can be determined accurately.

The computation of boundary crossing roots of the root-locus equation in (\ref{eq:rlocusK}) is a difficult problem. These roots and their crossing directions are computed in the following section.

\subsubsection{Computation of boundary crossing roots}
A root $s$  on the boundary of the root-locus region $\Re(s)=\so$ for the controller gain $\lambda$ satisfies the magnitude and phase equations of the root-locus equation in (\ref{eq:rlocusK}). We first find the intervals on the boundary where the magnitude condition holds for some $\lambda\in[0,\lambda_{\max}]$. This is equivalent to finding the intervals on $\Re(s)=\so$ and $\w\in[0,\infty)$ such that
\be \label{eq:Kcond}
\Lambda(\w):=h\so-\ln|G(\so+j\w)|\leq\ln \lambda_{\max}.
\ee
\begin{lem} \label{lem:polyK}
Assume that $G$ has neither poles nor zeros on the boundary of the root-locus region. The functions $\w \mapsto\Lambda(\w)$ and $\w\mapsto\Lambda'(\w)$ are continuous and the non-negative zeros of $\Lambda'(\w)$ are the non-negative real roots of the polynomial
\be \label{eq:polyKp}
\Gamma_z\sum_{i=1}^n\dwp\Gamma_p^i-\Gamma_{p}\sum_{r=1}^m\dwz\Gamma_z^r
\ee
where $\dsz=(\so-\sigma_{zr})$, $\dwz=(\w-\w_{zr})$, $\gamma_{zr}(\w)=\dsz^2+\dwz^2$, $\Gamma^r_z=\prod_{\substack{r_1=1\\r_1\neq r}}^m\gamma_{zr}(\w)$ for $r\in\N_1^m$, $\dsp=(\so-\sigma_{pi})$, $\dwp=(\w-\w_{pi})$, $\gamma_{pi}(\w)=\dsp^2+\dwp^2$, $\Gamma^i_p=\prod_{\substack{i_1=1\\i_1\neq k}}^n\gamma_{pi}(\w)$ for $i\in\N_1^n$, $\Gamma_z=\prod_{r=1}^m\gamma_{zr}(\w)$, $\Gamma_p=\prod_{i=1}^n\gamma_{pi}(\w)$.
\end{lem}

\noindent\textbf{Proof.\ }
Using the transfer function of G in (\ref{eq:tfG}), the function $\Lambda(\w)$ can be written as
\be \label{eq:Kw}
\Lambda(\omega)=h\so-\ln|\alpha|+\frac{1}{2}\left(\sum_{i=1}^n
\ln\gamma_{pi}(\omega)-\sum_{r=1}^m
\ln\gamma_{zr}(\omega)\right).
\ee
The first derivative of the function $\Lambda(\w)$ is
\be
\Lambda'(\w)=\sum_{i=1}^n \frac{\dwp}{\gamma_{pi}(\omega)}-\sum_{r=1}^m \frac{\dwz}{\gamma_{zr}(\omega)} \label{eq:Kwp}.
\ee
The functions $\Lambda(\w)$ and $\Lambda'(\w)$ are continuous except the points where $\gamma_{zr}(\w)$ or $\gamma_{pi}(\w)$ are equal to zero. These points are the poles or zeros of $G$ on $\Re(s)=\so$. The continuity results in Lemma \ref{lem:polyK} follow from the assumption. The polynomial given in (\ref{eq:polyKp}) is the numerator of the function $\Lambda'(\w)$ as in equation (\ref{eq:Kwp}) and the result follows.  \hfill $\Box$

\begin{cor} \label{cor:Kmonotonic}
Assume that $G$ has neither poles nor zeros on the boundary of the root-locus region. Then the function $\w\mapsto \Lambda(\w)$ is monotonic on the intervals whose boundary points are consecutive non-negative zeros of $\Lambda'(\w)$, $0$ and $\infty$.
\end{cor}
\noindent\textbf{Proof.\ }
By Lemma \ref{lem:polyK}, the function $\Lambda(\w)$ is continuous since $\so$ is chosen such that there are neither poles nor zeros of $G$ on $\Re(s)=\so$. Therefore it is monotonic inside the intervals determined by its extremum points and the end points of the boundary of the root-locus region, $0$ and $\infty$.~$\Box$

Since $\Lambda(\w)$ is monotonic on each interval defined in Corollary \ref{cor:Kmonotonic}, we find the subinterval in each interval where $\Lambda(\w)$ satisfies the inequality in (\ref{eq:Kcond}). This is done as follows. If both values of $\Lambda(\w)$ at the interval end points are smaller than $\ln \lambda_{\max}$, then all $\Lambda(\w)$ values in this interval are smaller than $\ln \lambda_{\max}$ because $\Lambda(\w)$ is monotonic. If one of the values of $\Lambda(\w)$ at the interval end points is larger and the other one is smaller than $\ln \lambda_{\max}$, we can find the point where $\Lambda(\w)$ is equal to $\ln \lambda_{\max}$ by root-finding algorithms for a bracketed root of monotonic continuous function (such as Brent's method \cite{BrentMethod}) and take the subinterval satisfying the inequality in (\ref{eq:Kcond}). If both values of $\Lambda(\w)$ at the interval end points are larger than $\ln \lambda_{\max}$, we discard that interval since all values of $\Lambda(\w)$ are larger than $\ln \lambda_{\max}$ and the inequality in (\ref{eq:Kcond}) never holds.

Based on this approach, we can compute the set of intervals $I$ on the boundary of the root-locus region where the magnitude condition (\ref{eq:Kcond}) is satisfied for some values of $\lambda\in[0,\lambda_{\max}]$.

The boundary crossing roots also satisfy the phase equation of the root-locus equation in (\ref{eq:rlocusK})
\be \label{eq:Phcond}
(2l+1)\pi=\phi(\w)\ \textrm{for} \   l\in\Z
\ee
over the intervals $I$ on $\Re(s)=\so$. Here the function $\phi(\w)$ represents the \emph{continuous} extension of the phase of the transfer function $G(s)e^{-hs}$. It satisfies the following equation,
\bd
\textrm{mod}\left(\angle \left.G(s)e^{-hs} \right|_{s=\so+j\w},2\pi\right)=\textrm{mod}\left(\phi(\w),2\pi\right)
\ed
where $\w\in[0,\infty)$. The left-hand side of equation (\ref{eq:Phcond}) represents constant functions of $\omega$. If we partition the intervals $I$ into the subintervals such that the function $\phi(\w)$ is monotonic on each subinterval, we can compute the boundary crossing roots by root-finding algorithms for a bracketed root of monotonic continuous function \cite{BrentMethod}. The following results allow us to compute the intervals on $\Re(s)=\so$ where the function $\phi(\w)$ is monotonic.

\begin{lem} \label{lem:polyPhi}
Assume that $G$ has neither poles nor zeros on the boundary of the root-locus region. Then the functions $\w\mapsto\phi(\w)$ and $\w\mapsto\phi'(\w)$ are continuous and the non-negative zeros of $\phi'(\w)$ are the non-negative real roots of the polynomial
\be
 \label{eq:polyPhi}
 \Gamma_{p}\sum_{r=1}^m\dsz\Gamma_z^r-\Gamma_z\sum_{i=1}^n\dsp\Gamma_p^i-h\Gamma_z\Gamma_{p}
 \ee
where the polynomials $\Gamma^r_z$ for $r\in\N_1^m$, $\Gamma^i_p$ for $i\in\N_1^n$, $\Gamma_z$ and $\Gamma_p$ are as defined in Lemma \ref{lem:polyK}.
\end{lem}
\noindent\textbf{Proof.\ }
Using the transfer function of $G$ in (\ref{eq:tfG}), the function $\phi(\w)$ is written as
\be \label{eq:Phi}
\phi(\w)=\phi_1(\w)+\phi_0
\ee where
\be \label{eq:Phi1}
\phi_1(\w):=\sum_{r=1}^m \tan^{-1} \frac{\dwz}{\dsz}
-\sum_{i=1}^n \tan^{-1}\frac{\dwp}{\dsp}
-h\w.
\ee The offset difference $\phi_0$ is defined as $\phi_0=\angle{G(\sigma_0)}-\phi_1(0)$ which is equal to $0$ or $\pi$.

The first derivative of the function $\phi(\w)$ is
\be \label{eq:Phip}
\phi'(\w)=\sum_{r=1}^m \frac{\dsz}{\gamma_{zr}(\w)}-
\sum_{i=1}^n \frac{\dsp}{\gamma_{pi}(\w)}-h.
\ee
Following the same arguments in Lemma \ref{lem:polyK}, the functions $\phi(\w)$ and $\phi'(\w)$ are continuous by the assumption. The polynomial given in (\ref{eq:polyPhi}) is the numerator of the function $\phi'(\w)$ as in equation (\ref{eq:Phip}) and the result follows. \hfill $\Box$

\begin{cor} \label{cor:polyPhimonotonic}
Assume that $G$ has neither poles nor zeros on the boundary of the root-locus region. Denote by  $I_\phi$ the set of intervals whose end points  are consecutive non-negative zeros of $\phi'(\w)$, $0$ and $\infty$. Then the function $\w\mapsto\phi(\w)$ is monotonic on each interval in the set $I_{\phi}$.
\end{cor}
\noindent\textbf{Proof.\ }
The function $\phi(\w)$ is continuous. The monotonicity of $\phi(\w)$ changes only at the points where $\phi'(\w)=0$. The assertion follows. \hfill $\Box$

The non-negative real roots of the polynomials (\ref{eq:polyKp}) and (\ref{eq:polyPhi}) are zeros of the rational functions (\ref{eq:Kwp}) and (\ref{eq:Phip}). These rational functions can be realized in a state-space representation as a series connection of second-order systems using the poles and zeros of $G$ and the time delay $h$ and the corresponding zeros can be computed by solving a generalized eigenvalue problem.

The intersection of two sets of intervals $I$ and $I_\phi$ partitions $I$ into the subintervals as  $I=\cup_{i=1}^{n_I}I_i$ where $\phi(\w)$ is monotonic on each interval $I_i$. Each intersection of the function $\phi(\w)$ and the constant functions in the left-hand side of equation (\ref{eq:Phcond}) over the intervals $I$ corresponds to a boundary crossing root since any such point on $I$ satisfies both the magnitude
condition in (\ref{eq:Kcond}) and the phase equation in (\ref{eq:Phcond}) of the root-locus equation on $\Re(s)=\so$. Since the function $\phi(\w)$ is monotonic on $I_i$, we can compute each intersection point by root-finding algorithms for a bracketed root of $\phi(\w)$ over the interval~$I_i$. The value of the $\w$ at the intersection point is the imaginary part of the boundary crossing root on the interval $I_i$ and the corresponding controller gain is the value of $e^{\Lambda(\w)}$ for this point. If there is no horizontal line intersecting $\phi(\w)$ on $I_i$, we discard this interval since there is no root crossing. Based on the above explanation, the following algorithm computes the boundary crossing roots of the root-locus equation in (\ref{eq:rlocusK}).

\begin{alg} \label{eq:bndxroot}
${}$\\
For each interval in $I_i=[\w_i^L,\w_i^R]$ of $I=\cup_{i=1}^{n_I}I_i$,
\begin{enumerate}
\item Compute the maximum and minimum values of the function $\phi$ over the interval $I_i$. Since the function is monotonic over the interval, they are the maximum and minimum values of the function $\phi$ at the interval ends points, i.e., $\phi_i^{\max}=\max\{\phi(\w_i^L),\phi(\w_i^R)\}$ and $\phi_i^{\min}=\min\{\phi(\w_i^L),\phi(\w_i^R)\}$.
\item Consider the integers $l$ for which the constant function $y=(2l+1)\pi$ has an intersection with the function $y=\phi(\w)$ over the interval $I_i$. Note that the integers satisfy $l\in[l_i^{\min},l_i^{\max}]$ where $l_i^{\min}=\left\lceil \frac{\phi_i^{\min}}{2\pi}-\frac{1}{2} \right\rceil$
     and $l_i^{\max}=\left\lfloor \frac{\phi_i^{\max}}{2\pi}-\frac{1}{2} \right\rfloor$.
\item If $(l_i^{\min}>l_i^{\max})$ \\
discard the interval $I_i$, \\
else \\
for each integer $l$ from $l_i^{\min}$ to $l_i^{\max}$
\begin{itemize}
\item compute the intersection point between the horizontal line $(2l+1)\pi$ and the function $\phi$ over the interval $I_i$. This point can be considered as a the bracketed root of a monotonic continuous function $\phi$ over $I_i$ and accurately computed by root-finding algorithms such as Brent's method \cite{BrentMethod}. Denote this point by $\w_{cr}$ which is equal to the imaginary part of the boundary crossing root.
\item compute the corresponding controller gain for the boundary crossing root, i.e., $\lambda_{cr}=e^{\Lambda(\w_{cr})}$.
\end{itemize}
\end{enumerate}
\end{alg}

By Algorithm~\ref{eq:bndxroot}, we compute all boundary crossing roots of the root-locus equation in (\ref{eq:rlocusK}) and their corresponding controller gain values for $\lambda\in[0,\lambda_{\max}]$. Their crossing directions are determined by the following theorem.

\begin{thm} \label{thm:CD}
The crossing direction of a boundary crossing root $s_{cr}=\so+j\w_{cr}$ only depends on the imaginary part $\w_{cr}$ on the boundary and is equal to
\be
\mathcal{CD}(s_{cr},\lambda_{cr})=-\textrm{sgn}\left(\phi'(\w_{cr})\right).
\ee
\end{thm}
\noindent\textbf{Proof.\ }
Using the transfer function $G$ in (\ref{eq:tfG}) and (\ref{eq:Kwp},\ref{eq:Phip}), we obtain
\be \label{eq:GpG}
G'(s_{cr})G^{-1}(s_{cr})-h=\phi'(\w_{cr})+j \Lambda'(\w_{cr}).
\ee
 By the crossing direction formula in (\ref{eq:RT}) and the equation in (\ref{eq:GpG}), the crossing direction of $s_{cr}$ at $\lambda=\lambda_{cr}$ is equal to
\bd
\mathcal{CD}(s_{cr},\lambda_{cr})=\textrm{sgn}\left(\Re\left(\left(\lambda_{cr}
\left(h-\frac{G'(s_{cr})}{G(s_{cr})}\right)\right)^{-1}\right)\right)
\ed
\bd
\hspace{2.0cm} =-\textrm{sgn}(\phi'(\w_{cr})). \hspace*{3.4cm} \Box
\ed
By Theorem \ref{thm:CD}, the crossing directions of the roots are the same when their imaginary parts are inside the same interval of $I_\phi$ (see Corollary \ref{cor:polyPhimonotonic}). We determine the crossing directions of boundary crossing roots from their imaginary parts. We group them according to their crossing directions and define the sets $W^{in}=\{\ s_\nu^I,\ \lambda_\nu^I\ \}_{\nu=1}^{n_i}$ and $W^{out}=\{\ s_\nu^O,\ \lambda_\nu^O\ \}_{\nu=1}^{n_o}$ where $s_\nu^I=\so+j\w_\nu^I,\ \lambda_\nu^I$ for $\nu\in\N_1^{n_i}$ and $s_\nu^O=\so+j\w_\nu^O,\ \lambda_\nu^O$ for $\nu\in\N_1^{n_o}$ are the boundary crossing roots entering into or leaving the root-locus region and their controller gains respectively.

\noindent\textbf{Remark 1:\ } The crossing direction formula in (\ref{eq:RT}) is well-posed (either $+1$ or $-1$) if there are neither poles nor zeros of $G$ or branch points on the boundary of the root-locus region.\\
\noindent\textbf{Remark 2:\ } The closed-loop system is of \emph{neutral} type (see, e.g., \cite{WimBook}) when $G$ is biproper (i.e., $d:=G(\infty)\neq0$). Its asymptotic root chains lie outside the root-locus region $\Cs$ for $\lambda_{\max}<\frac{e^{h\so}}{|d|}$, while the region always has infinitely many roots for larger controller gains. Since it is not numerically possible to follow infinitely many roots, $\lambda_{\max}$ is assumed to be smaller than $\frac{e^{h\so}}{|d|}$.

\subsection{Time delay as locus parameter} \label{sec:critptsH}
The root-locus equation w.r.t. the time delay is given in (\ref{eq:rlocusH}). The starting points of the root locus are the zeros of the transfer function $1+G(s)$ inside the root-locus region $\Cs$.

The branch points satisfy the root-locus equation (\ref{eq:rlocusH}) and
\begin{equation}\label{poleq}
\frac{\partial f(s,\lambda)}{\partial s}=(G^{\prime}(s)-G(s)\lambda)e^{-\lambda s}=0.
\end{equation}
Since the locus parameter $\lambda=h$ in equation (\ref{poleq}) is unknown, the direct computation of branch points is difficult compared to the previous case. Therefore we detect the branch points of the root locus while following the root-locus trajectory as described in Section \ref{ssec:detectbranch}.

The relative stability analysis of the closed-loop of SISO dead-time systems w.r.t. the stability boundary $\Re(s)=\so$ is given in \cite{GumussoyTIMC2010}. This analysis is based on the computation of boundary crossing roots of the root-locus equation in (\ref{eq:rlocusH}) which is similar to the computation for the controller gain case in Section~\ref{sec:critptsK}. Further information can be found in \cite{GumussoyTIMC2010}. Note that when the plant is  biproper, $\lambda_{\max}$ is assumed to be smaller than $\max\left(0,\frac{\ln |d|}{|\sigma_0|}\right)$ to have finitely many characteristic roots inside the root-locus region $\Cs$ (see Remarks $2$ and $3$ in \cite{GumussoyTIMC2010}).

\section{Computing a Root-Locus Trajectory} \label{sec:rloci}
We compute the root-locus trajectory of a characteristic root between two critical points. The trajectory satisfies the root-locus equation w.r.t the controller gain in (\ref{eq:rlocusK}) or the time delay in (\ref{eq:rlocusH}), represented by the locus parameter $\lambda$.

The starting points and the characteristic roots entering into the root-locus region are computed in Section~\ref{sec:critpts}. We follow each root-locus trajectory by a secant-predictor, Newton-corrector continuation method \cite{ContinuationSIAMBook}. In the prediction step, a line passing through the last two computed roots and locus parameters is used to estimate the next root and the next locus parameter at a certain distance (step length) in the (characteristic root, locus parameter)-space. This estimate is corrected using Newton's method in the correction step. The next iteration continues in a similar way, though the step length is adaptive. The branch points for the controller gain are computed apriori and those for the time delay are detected while following the trajectory. The new branch direction is determined by Lemma~\ref{lem:branch} and the trajectory is followed until the upper bound of the locus parameter $\lambda_{\max}$ is reached.

\subsection{Parameterization of the root-locus trajectory}
Continuation based methods follow the characteristic root $s$ and the locus parameter $\lambda$ based on the parameterization of the root-locus trajectory. In the literature, the characteristic root is parameterized w.r.t. the locus parameter as $s=s(\lambda)$ for computational purposes~\cite{FioravantiTDS2010,Suh82}. In this parameterization, the characteristic roots are highly sensitive w.r.t. changes of the locus parameter in the neighborhood of branch points and relevant numerical problems are reported in \cite{FioravantiTDS2010}. We illustrate this sensitivity on the root locus  by means of equation (\ref{eq:rlocusK}), where $h=\frac{4\pi}{3}$ and $G(s)=\frac{s^2}{(s^2+1)(s^2+1)}$.

\begin{figure}[!h]
\centering
        \includegraphics[width=0.45\textwidth]{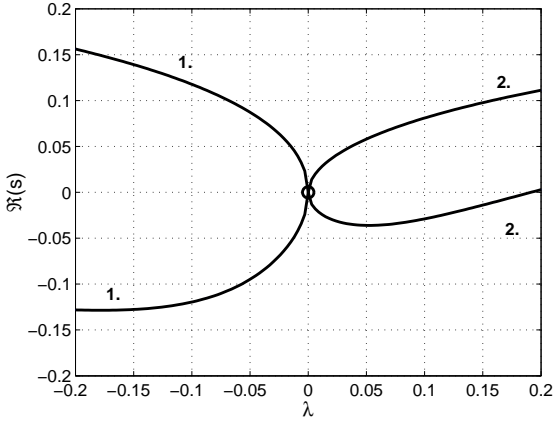}
\caption{\label{fig:param} The rightmost characteristic roots of the root-locus equation in~(\ref{eq:rlocusK}) for the plant with $h=\frac{4\pi}{3}$ and $G(s)=\frac{s^2}{(s^2+1)(s^2+1)}$ as a function of the controller gain $\lambda$.}
\end{figure}

Figure~\ref{fig:param} shows the high sensitivity of the characteristic roots w.r.t. changes of $\lambda$ at $\lambda=0$ where a branch point occurs. As can be seen in Figure~\ref{fig:param}, this branch point  appears as a turning point in the root versus locus-parameter space. When a continuation method tries to follow the trajectory $1$ or $2$, the turning point around $\lambda=0$ causes several numerical problems if $\lambda$ is used as a continuation parameter. Take for instance trajectory 1. First the sensitivity or derivative of the root w.r.t.~$\lambda$ tends to infinity when $\lambda\rightarrow 0-$. Second when the $\lambda$ passes zero the same branch cannot be followed anymore because the original trajectory turns back. In the best scenario a point on trajectory 2~will be computed but this is unlikely with a local method because trajectory 2 has another direction than trajectory 1 (see Lemma~\ref{lem:branch}), hence the estimate of the root obtained from the previous points will be bad. On the other hand, trajectory 1 and trajectory 2 appear as smooth  curves in the complex plane as shown in Figure~\ref{fig:param}. In the field of numerical bifurcation analysis (see, e.g., \cite{Seydel1994}), this was the motivation to parameterize the trajectories in terms of a natural notion of an arclength, i.e., a distance measured along the trajectories.

\begin{figure}[!h]
\centering
        \includegraphics[width=0.45\textwidth]{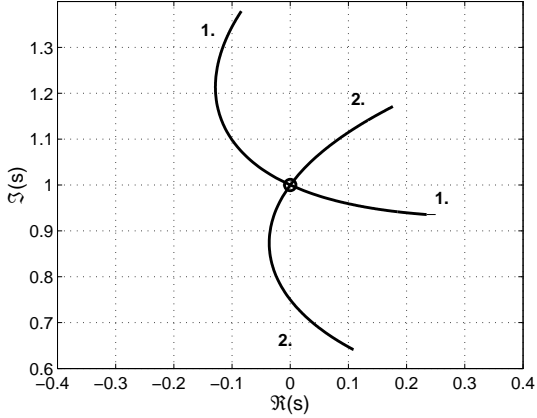}
\caption{The root-locus trajectories $1$ and $2$ in Figure~\ref{fig:param} on the complex plane.}
\end{figure}

\begin{figure}[!h]
\centering
        \includegraphics[width=0.45\textwidth]{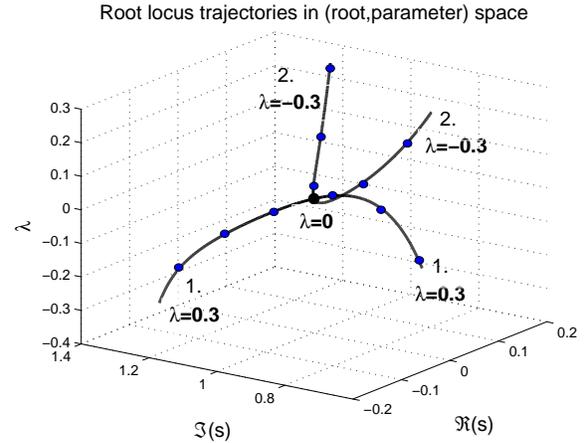}
\caption{\label{fig:3dparam} The root-locus trajectories $1$ and $2$ in Figure~\ref{fig:param} in the (characteristic root, locus parameter)-space.}
\end{figure}

In our method we parameterize the root-locus trajectories in terms of the arclength $\gamma$ measured along the trajectory in the combined (characteristic root $s$, locus parameter $\lambda$) space, i.e., $s=s(\gamma)$ and $\lambda=\lambda(\gamma)$ for $\gamma>0$.  For the example given in (\ref{ex1}) the two root-locus trajectories $1$ and $2$ are shown in Figure~\ref{fig:3dparam}. The trajectories are smooth and using equal arclength steps for $\gamma$ (where each step is shown as a point on the trajectory) the trajectories are traced without any numerical problem. Figure~\ref{fig:3dparam} also illustrates that a branch point is characterized by an extremum of the function $\gamma\mapsto\lambda(\gamma)$ which can be used to detect the presence of a branch point along the trajectory.

\subsection{Prediction step}
\begin{figure}[!h]
\centering
        \includegraphics[width=0.40\textwidth]{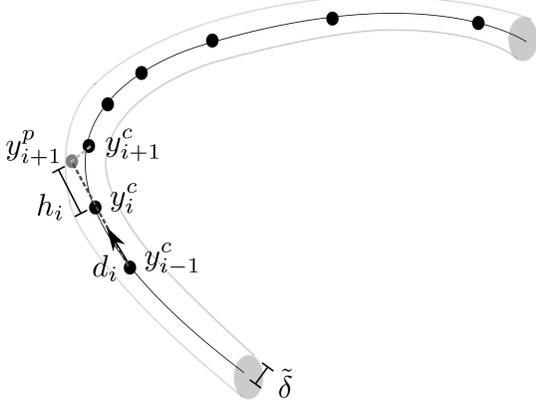}
\caption{\label{fig:rlocustraj} Computing the root-locus trajectory: Prediction, correction steps and adaptively chosen step lengths.}
\end{figure}

The predicted root and the locus parameter computation in the prediction step requires the previous root, the locus parameter, a direction and a step length. For each root-locus trajectory, the starting point $s_0$ is available. The direction of the prediction step $d_i$ is computed as follows:
\begin{itemize}
\item The initial directions $d_0^{*}\in\C$ for the poles of $G$ inside $\Cs$ and the boundary crossing roots are computed by the root-locus equation, i.e., by solving
    \begin{equation}\label{extra1}
    \left.\left[
        \begin{array}{cc}
        \frac{\partial f}{\partial s}\ & \frac{\partial f}{\partial \lambda}
        \end{array}
    \right]
    \right|_{\substack{
           f(s,\lambda)=0
            }}
    d_0^*=0
    \end{equation}
where $d_0^*=[d_0^s\ \ d_0^\lambda]^T$ is normalized. Set the root-locus direction as $d_i=[\Re(d_0^s) \ \Im(d_0^s) \ d_0^\lambda]^T$.
                                          Note that (\ref{extra1}) is the mathematical characterization of the \emph{tangent} vector to the trajectory in the $(s,\lambda)$ space~\cite{Seydel1994}.
\item The directions in other iterations are computed using the real and imaginary parts of last two corrected roots and locus parameters,
$y_i^c=[\sigma_i^c\  \w_i^c\  \lambda_i^c]^T$ and
$y_{i-1}^c=[\sigma_{i-1}^c\ \w_{i-1}^c\ \lambda_{i-1}^c]^T$ by
\be \label{eq:preddirection}
d_i=\frac{y_i^c-y_{i-1}^c}{\|y_i^c-y_{i-1}^c\|}\ \textrm{for}\ i\geq 1.
\ee
\end{itemize}
The real and imaginary parts of the predicted root and the locus parameter $y_{i+1}^p=[\sigma_{i+1}^p\ \w_{i+1}^p \ \lambda_{i+1}^p)]^T$
are computed using a line equation with a step length $h_i$
\be \label{eq:secpred}
y_{i+1}^p=y_i^c+d_i h_i\ \textrm{for}\ i\geq 0.
\ee The geometric illustration of the prediction step is given in Figure~\ref{fig:rlocustraj} where the predicted point $y_{i+1}^p$ is shown as a gray dot. The initial step length $h_0$ is fixed. The step lengths in other iterations are calculated adaptively based on previous values as we outline later on.

\subsection{Correction step}
We use Newton's method to solve a set of nonlinear equations to find the real and imaginary parts of the corrected root and the locus parameter $y_{i+1}^c=[\sigma_{i+1}^c\ \w_{i+1}^c\ \lambda_{i+1}^c]^T$. These equations are given by
\begin{align}
M(\sigma_{i+1}^c,\w_{i+1}^c,\lambda_{i+1}^c)&=0, \label{eq:Mnewton}\\
P(\sigma_{i+1}^c,\w_{i+1}^c,\lambda_{i+1}^c)&=0, \label{eq:Nnewton} \\
(y_{i+1}^c-y_{i+1}^p) d_i&=0. \label{eq:line}
\end{align}
The functions $M$ and $P$ are equivalent representations of the magnitude and phase equations of the root-locus equation. Equation (\ref{eq:line}) guarantees that the linearized distance of the corrected root and the locus parameter $y_{i+1}^c$ from the predicted root and the locus parameter $y_{i+1}^p$ is equal to the step size $h_i$.  Because of the linearization the proposed method can be seen as a \emph{pseudo arclenghth continuation} method~\cite{Seydel1994}.

The function $M$ and $P$ are defined using the root-locus equation w.r.t. the controller gain in (\ref{eq:rlocusK}) or the time delay in (\ref{eq:rlocusH}) as \vspace{-0.8cm}
{\small
\bd
M(\sigma,\w,\lambda)=
\left\{
\begin{array}{c}
\mathcal{M}(\sigma,\w,\lambda,h) \\
\mathcal{M}(\sigma,\w,1,\lambda)
\end{array},
\right.
P(\sigma,\w,\lambda)=
\left\{
\begin{array}{c}
\mathcal{P}(\sigma,\w,\lambda,h) \\
\mathcal{P}(\sigma,\w,1,\lambda)
\end{array}
\right.
\ed} where \vspace{-1cm}
{\small
\bea
\nonumber \mathcal{M}(\sigma,\w,k,h)&=&\ln|\alpha|+\frac{1}{2}\sum_{r=1}^m \left(\ln (\sigma-\sigma_{zr})^2+(\w-\w_{zr})^2\right) \\
 &&\hspace{-1.5cm} -\frac{1}{2}\sum_{i=1}^n \left(\ln (\sigma-\sigma_{pi})^2+(\w-\w_{pi})^2\right)-h\sigma+\ln k, \label{eq:M}\\
\nonumber \mathcal{P}(\sigma,\w,k,h)&=&\angle \alpha+\sum_{r=1}^m \tan^{-1}\frac{\w-\w_{zr}}{\sigma-\sigma_{zr}} \\
&& \hspace{1cm}   -\sum_{i=1}^n \tan^{-1}\frac{\w-\w_{pi}}{\sigma-\sigma_{pi}}-h\w-\pi. \label{eq:P}
\eea}Note that the controller gain $k$ and the time delay $h$ have constant values in the root-locus equation w.r.t. the time delay in (\ref{eq:rlocusH}) and the controller gain in (\ref{eq:rlocusK}) respectively.

The function $P(\sigma,\w,\lambda)$ has a range $(-\pi,\pi]$ and the arctangent functions in $P(\sigma,\w,\lambda)$ have ranges $(-\pi,\pi]$ (similarly to the two argument function atan2 in MATLAB).

We compute the corrected point $y_{i+1}^c$ by Newton's method on (\ref{eq:Mnewton}-\ref{eq:line}), i.e., by iteratively solving
\be
J_{i+1}^k(\tilde{y}_{i+1}^{k+1}-\tilde{y}_{i+1}^{k}) =f_{i+1}^k,\ \textrm{for}\ k=0,1,\ldots
\ee where $J_{i+1}^k$ and $f_{i+1}^k$ are the Jacobian and the gradient vector evaluated at  $\tilde{y}_{i+1}^{k}$. The iteration is initialized by the predicted point such that $\tilde{y}_{i+1}^0=y_{i+1}^p$. We stop the iteration when the prescribed accuracy is reached and set the corrected point $y_{i+1}^c$ to the last point of the iteration. The geometric illustration is given in Figure~\ref{fig:rlocustraj}.

Regarding the computational complexity, the above computation requires evaluating the functions on the left-hand side of the equations in (\ref{eq:Mnewton}-\ref{eq:line}) and the corresponding Jacobian and gradient of these equations at each iteration. This computation is numerically cheap and uses only the poles and zeros of $G$ and the time delay $h$.



\subsection{Adaptive step length}
The step length computation for the next prediction step depends on two factors \cite{ContinuationSIAMBook}
\begin{itemize}
\item the \emph{contraction rate} of the first two consecutive Newton steps in the corrector step, i.e.,
$
\kappa_{i+1}:=\frac{\left\|(J_{i+1}^0)^+ f_{i+1}^{1}\right\|}{\left\|(J_{i+1}^0)^+ f_{i+1}^{0}\right\|};
$
\item the \emph{distance} to the root-locus trajectory \\
\\
$
\delta_{i+1}=\left\|1-e^{M(\sigma_{i+1}^c,\w_{i+1}^c,\lambda_{i+1}^c)+jP(\sigma_{i+1}^c,\w_{i+1}^c,\lambda_{i+1}^c)}\right\|.
$
\end{itemize}
The individual deceleration factors are calculated as $\kappa_{df}=\sqrt{\frac{\kappa_{i+1}}{\tilde{\kappa}}}$ and $\delta_{df}=\sqrt{\frac{\delta_{i+1}}{\tilde{\delta}}}$ where $\tilde{\kappa}$ and $\tilde{\delta}$ are the nominal contraction rate and the distance. The overall deceleration factor $h_{df}$ of the step length is the maximum of individual deceleration factors, $\kappa_{df}$ and $\delta_{df}$ limited to $[\frac{1}{2},2]$, i.e., $h_{df}:=\max\{\min\{\max\{\kappa_{df}, \delta_{df}\},2\},\frac{1}{2}\}$.

Note that if $h_{df}=2$, the predictor step is repeated with a reduced step length. This check is done inside the corrector step to avoid unnecessary Newton iterations (see \cite{ContinuationSIAMBook} for further details). The step length for the next prediction step is $h_{i+1}=h_i/h_{df}$.

The adaptive step length selection allows us to follow the root-locus trajectory in a computationally efficient way with a prescribed accuracy. When the trajectory has sharp (resp. wide) curves, the step lengths are smaller (resp. larger). This behavior is illustrated in Figure~\ref{fig:rlocustraj}.


\subsection{Detection of branch points} \label{ssec:detectbranch}

We directly computed the branch points of the root locus w.r.t. the controller gain in Section~\ref{sec:critptsK}. The branch point detection of the root locus w.r.t. the time delay requires monitoring the time delay parameter along the root-locus trajectory. By Lemma~\ref{lem:branch}, the root-locus trajectory passes through the branch point when the multiplicity of the branch point is odd. In this case, there is no need to detect the branch point. Note that the locus parameter is strictly increasing for this case.

When the multiplicity of the branch point is even, the continuation method follows another root-locus trajectory after passing the branch point. The locus parameter on this trajectory is strictly decreasing since we follow the branch in the opposite direction. When we detect that the current time delay value is smaller than previous time delay values, we know that we have passed a branch point. Since we bracketed the branch point by the current and previous points, we then solve a set of nonlinear equations characterizing the branch point. Based on the multiplicity of the branch point, we decide the branch direction by Lemma~\ref{lem:branch}. The branch point and its direction are added into the list for the next trajectories.

\section{Algorithm} \label{sec:alg}
The overall algorithm is as follows.
\begin{enumerate}
\item Compute the critical points of the root locus.
\item For each root-locus trajectory:
\begin{enumerate}
\item Starting with the initial point, find the next point by computing the predicted point $y_{i+1}^p$, the corrected point $y_{i+1}^c$ and update the step length for the next step $h_{i+1}$.
\item Continue to compute the next point until one of the following conditions holds:
\begin{enumerate}
\item The trajectory reaches a branch point. Go to Step $2-a)$ and compute the next point using the new direction by Lemma~\ref{lem:branch}.
\item The locus parameter exceeds $\lambda_{\max}$ or the trajectory leaves $\Cs$. Stop the computation for this trajectory. Update the initial point with the starting point of a new root-locus trajectory and Go to Step $2-a)$.
\end{enumerate}
\end{enumerate}
\item Stop if there is no remaining root-locus trajectory.
\end{enumerate}
Note that due to the symmetry of the spectrum, on the real axis it is sufficient to compute the branch points and continue from the next branch point, i.e., a continuation method is not needed. In our implementation, we set the nominal contraction rate, the nominal distance and the tolerance for the corrector step to $\tilde{\kappa}=0.5$, $\tilde{\delta}=10^{-3}$ and $10^{-5}$.
%
\section{Examples} \label{sec:ex}
\begin{figure}[!h]
\begin{center}
\includegraphics[width=0.45\textwidth]{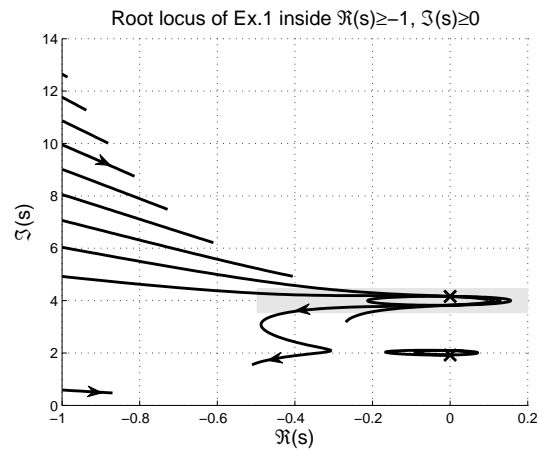}
\caption{\label{fig:rlocus1} The root locus of Example \ref{ex1} for time delay $\lambda\in[0,5]$ inside $\Re(s)\geq-1$, $\Im(s)\geq0$.}
\end{center}
\end{figure}

\begin{figure}[!h]
\begin{center}
\includegraphics[width=0.45\textwidth]{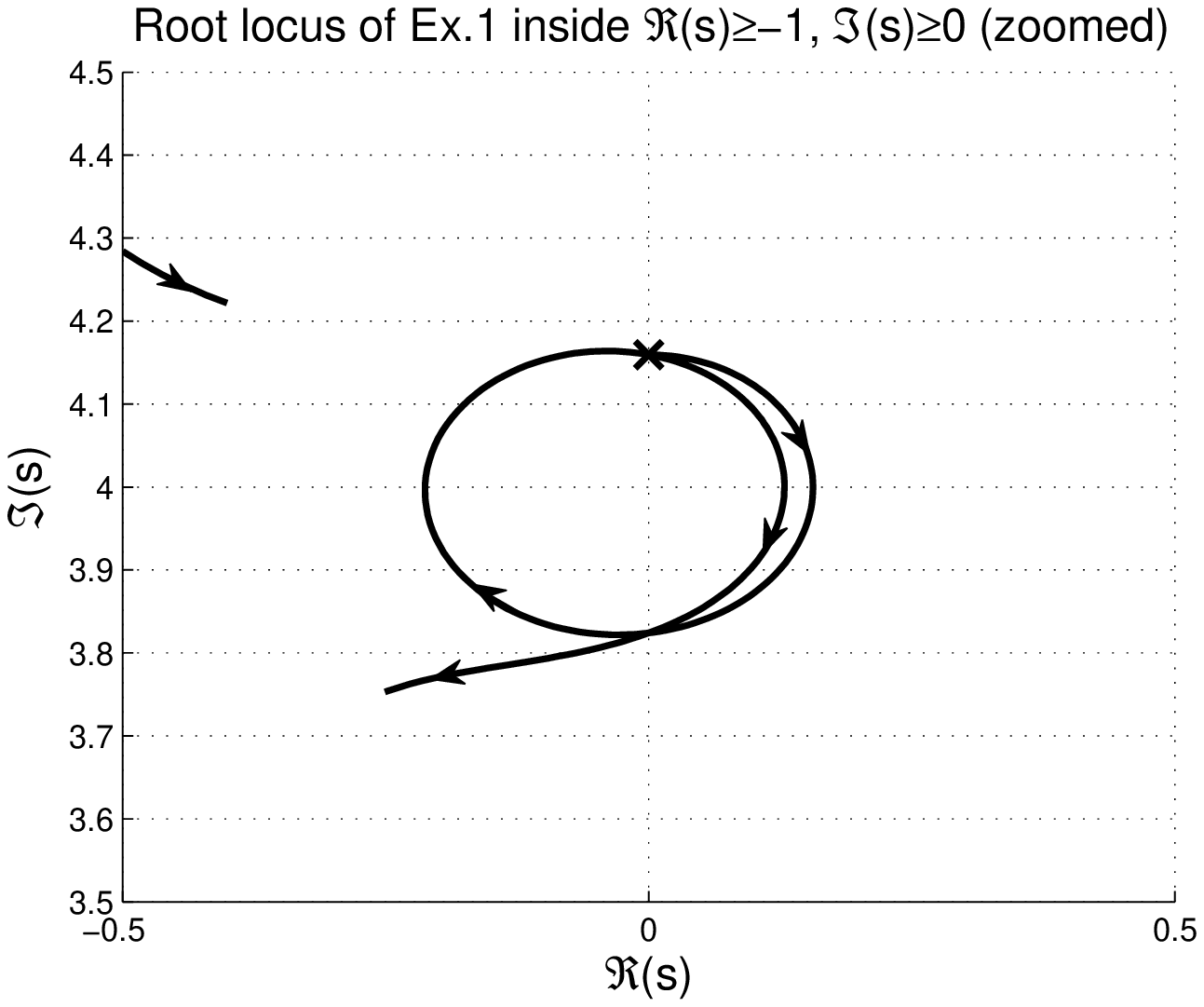}
\caption{\label{fig:rlocus1a} The root locus inside the gray region in Figure~\ref{fig:rlocus1} for time delay $\lambda\in[0,2.7]$.}
\end{center}
\end{figure}

\begin{figure}[!h]
\begin{center}
\includegraphics[width=0.45\textwidth]{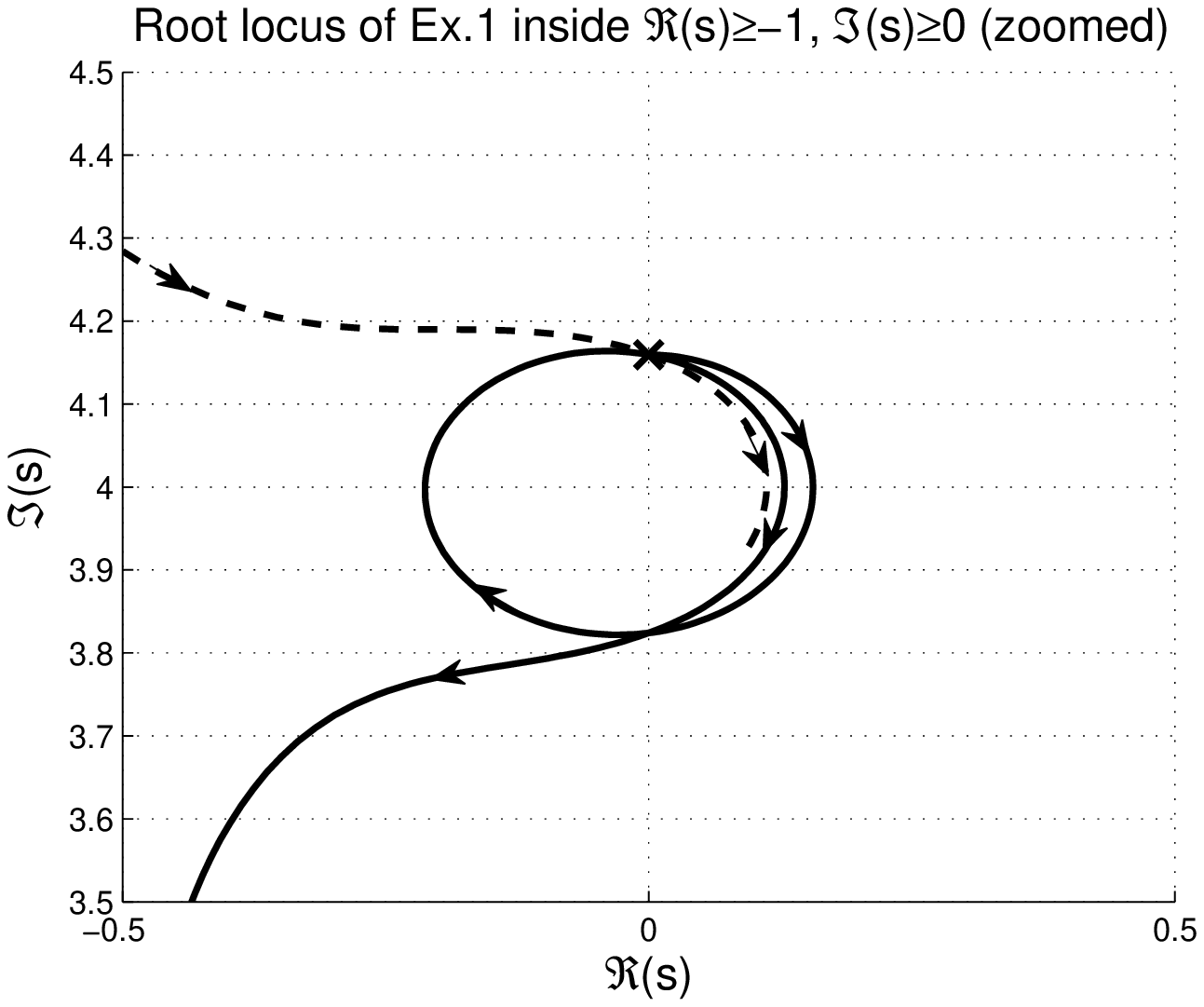}
\caption{\label{fig:rlocus1b} The root locus inside the gray region in Figure~\ref{fig:rlocus1} for time delay  $\lambda\in[0,3.8]$.}
\end{center}
\end{figure}

\begin{figure}[!h]
\begin{center}
\includegraphics[width=0.45\textwidth]{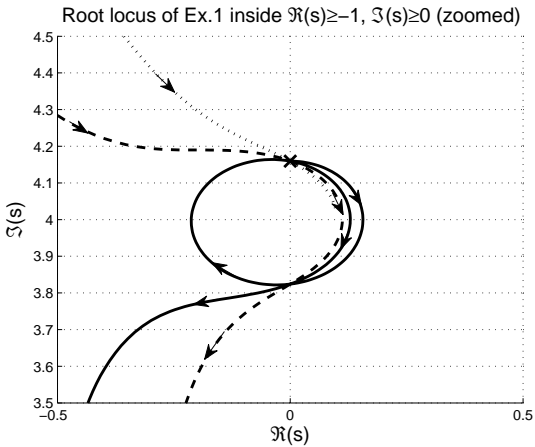}
\caption{\label{fig:rlocus1c} The root locus inside the gray region in Figure~\ref{fig:rlocus1} for time delay $\lambda\in[0,5]$.}
\end{center}
\end{figure}

We consider three examples with unique characteristics on their root locus. Example \ref{ex1} has circular trajectories close to each other. Trajectories of Example \ref{ex2} require different order of accuracy. Example~\ref{ex3} has a branch point, a trajectory going out of the region and trajectories with large complex numbers. We show root-locus plots only on the complex upper half-plane since the root-locus plots are symmetric w.r.t. the real axis due to real-valued coefficients of the plants in Examples.
\begin{exmp} \label{ex1}
In \cite{MichielsJVC10}, the stability of equation (\ref{eq:rlocusH}) is considered for the oscillator system, $G(s)=\frac{\epsilon s^2}{(s^2+\w_1^2)(s^2+\w_2^2)}$. The behavior of the rightmost characteristic roots is analyzed as a function of the delay parameter for small values of a gain parameter. In a particular example, the parameters of the system are set to $\epsilon=1$, $\w_1=2$, $\w_2=4$ and the oscillatory nature of the rightmost characteristic root and the stability of the closed-loop system are shown.

We plot the root locus in equation (\ref{eq:rlocusH}) of this plant inside $\Re(s)\geq-1$ for $\lambda\in[0,5]$ in Figure~\ref{fig:rlocus1}. It is well-known that the characteristic roots of time delay systems can only cross the imaginary axis at a finite number of points when the delay is varied. This property can be clearly seen in Figures~\ref{fig:rlocus1}-\ref{fig:rlocus1c}. Note in particular in Figure~\ref{fig:rlocus1c}  how characteristic roots are passing through same points on the imaginary axis. Root-locus plots in Figures~\ref{fig:rlocus1a}-\ref{fig:rlocus1c} illustrate the oscillatory behavior of the characteristic root trajectories centered around $\pm2j$, $\pm4j$ and the asymptotic characteristic roots coming from the left. This phenomenon is qualitatively explained in~\cite{MichielsJVC10}. The closed-loop system is stable for the time delay $\lambda\in[0.83,1.50]\cup[4.11,4.50]$.
\end{exmp}

\begin{figure}[!h]
\begin{center}
\includegraphics[width=0.45\textwidth]{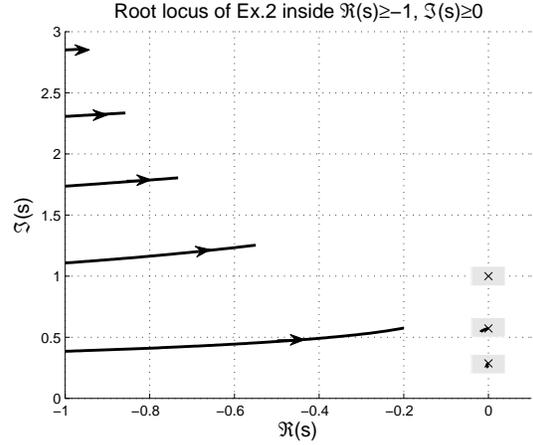}
\caption{\label{fig:rlocus2} The root locus of Example \ref{ex2} for controller gain $\lambda\in[0,6]$ inside $\Re(s)\geq-1$, $\Im(s)\geq0$.}
\end{center}
\end{figure}

\begin{figure}[!h]
\begin{center}
\includegraphics[width=0.45\textwidth]{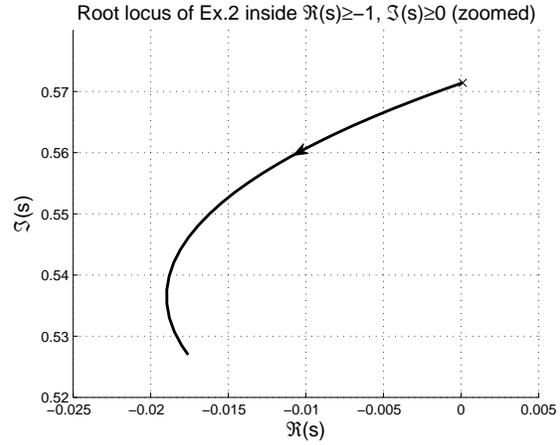}
\caption{\label{fig:rlocus2a} The root locus inside the middle gray region in Figure~\ref{fig:rlocus2} for controller gain $\lambda\in[0,6]$.}
\end{center}
\end{figure}

\begin{figure}[!h]
\begin{center}
\includegraphics[width=0.45\textwidth]{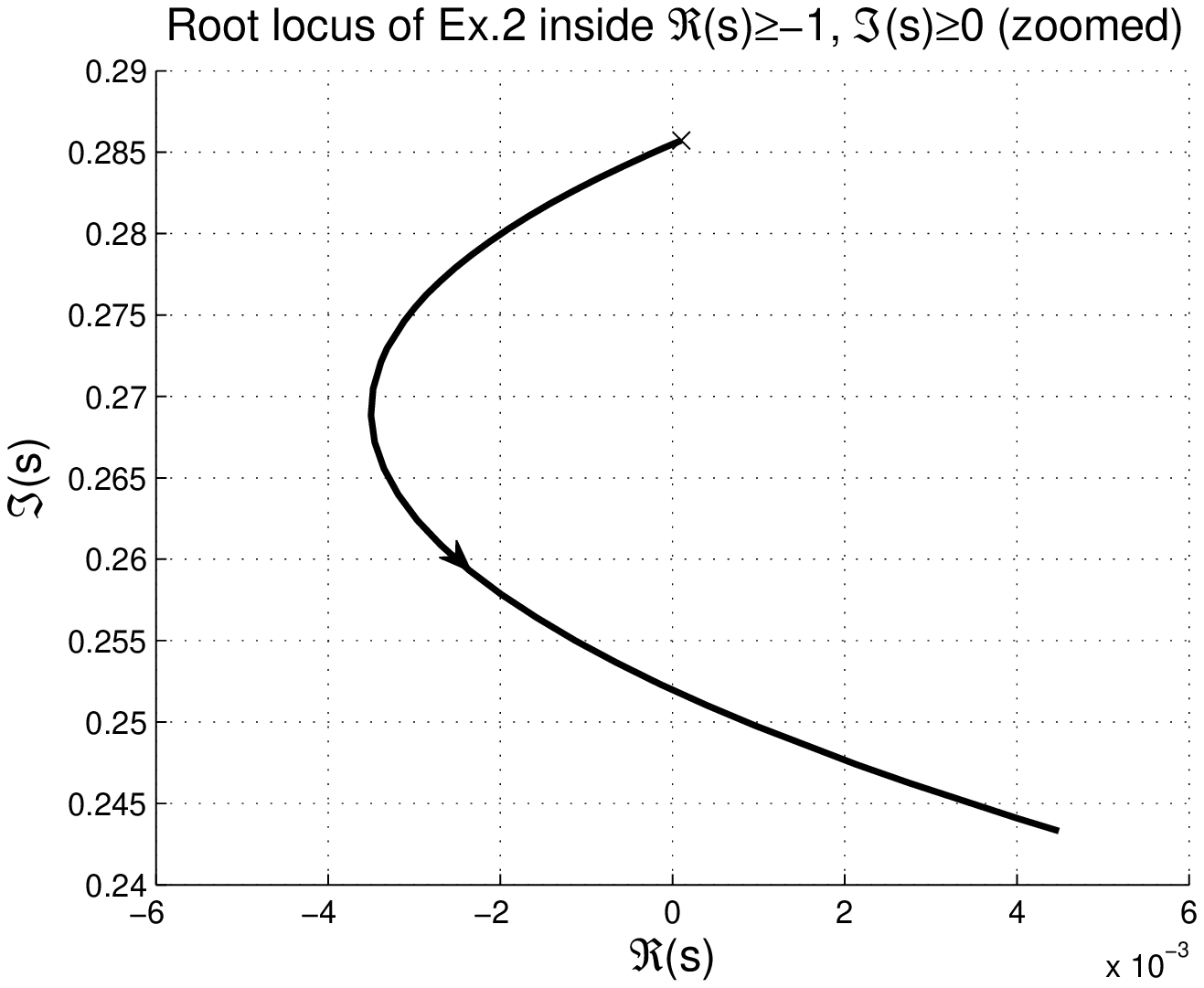}
\caption{\label{fig:rlocus2b} The root locus inside the bottom gray region in Figure~\ref{fig:rlocus2} for controller gain $\lambda\in[0,6]$.}
\end{center}
\end{figure}

\begin{figure}[!h]
\begin{center}
\includegraphics[width=0.45\textwidth]{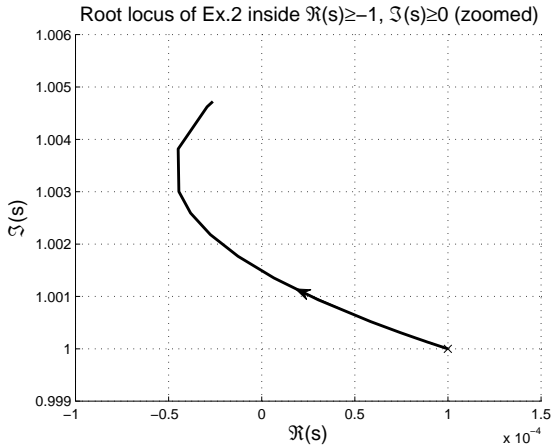}
\caption{\label{fig:rlocus2c} The root locus inside the top gray region in Figure~\ref{fig:rlocus2} for controller gain $\lambda\in[0,6]$.}
\end{center}
\end{figure}

\begin{exmp} \label{ex2}
The following SISO dead-time system is borrowed from \cite{ComplexTDSBook}, p. $80$-$81$ where $h=12.48$ and \vspace{-1cm}
{\small
\begin{multline*}
G(s)=10^{-3}\left( s^6-6\  10^{-4}s^5+1.4081634\ s^4 \right.\\
-5.6326533\ 10^{-4}\ s^3+0.43481891\ s^2-8.6963771\ 10^{-5}\ s \\
\left.+2.6655565\ 10^{-2}\right)^{-1}.
\end{multline*}
} The root-locus plot for equation (\ref{eq:rlocusK}) of this plant is given in Figure~\ref{fig:rlocus2}. As shown with cross markers inside gray regions in the figure, the plant has $6$ unstable poles. They have all moved to the complex left half-plane at $\lambda=1.860$ and the closed-loop system becomes stable for the controller gain  $\lambda\in[1.860,4.469]$. Note that the root-locus trajectories have lengths in the order of magnitudes, $1$, $10^{-2}$, $10^{-3}$, $10^{-4}$ respectively as illustrated in Figures~\ref{fig:rlocus2}-\ref{fig:rlocus2c}. We follow each trajectory without any problem due to the adaptive step length in the algorithm.
\end{exmp}

\begin{figure}[!h]
\begin{center}
\includegraphics[width=0.45\textwidth]{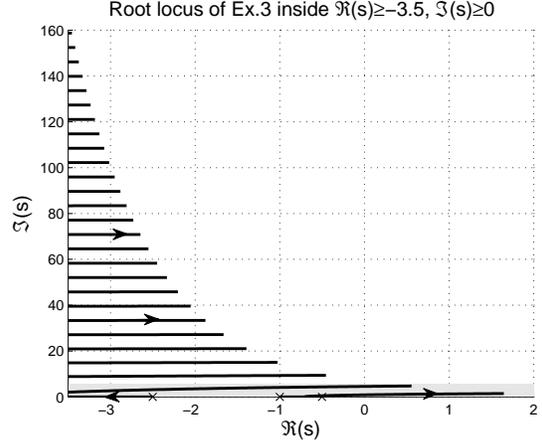}
\caption{\label{fig:rlocus3} The root locus of Example \ref{ex3} for controller gain $\lambda\in[0,5]$ inside $\Re(s)\geq-3.5$, $\Im(s)\geq0$.}
\end{center}
\end{figure}

\begin{figure}[!h]
\begin{center}
\includegraphics[width=0.45\textwidth]{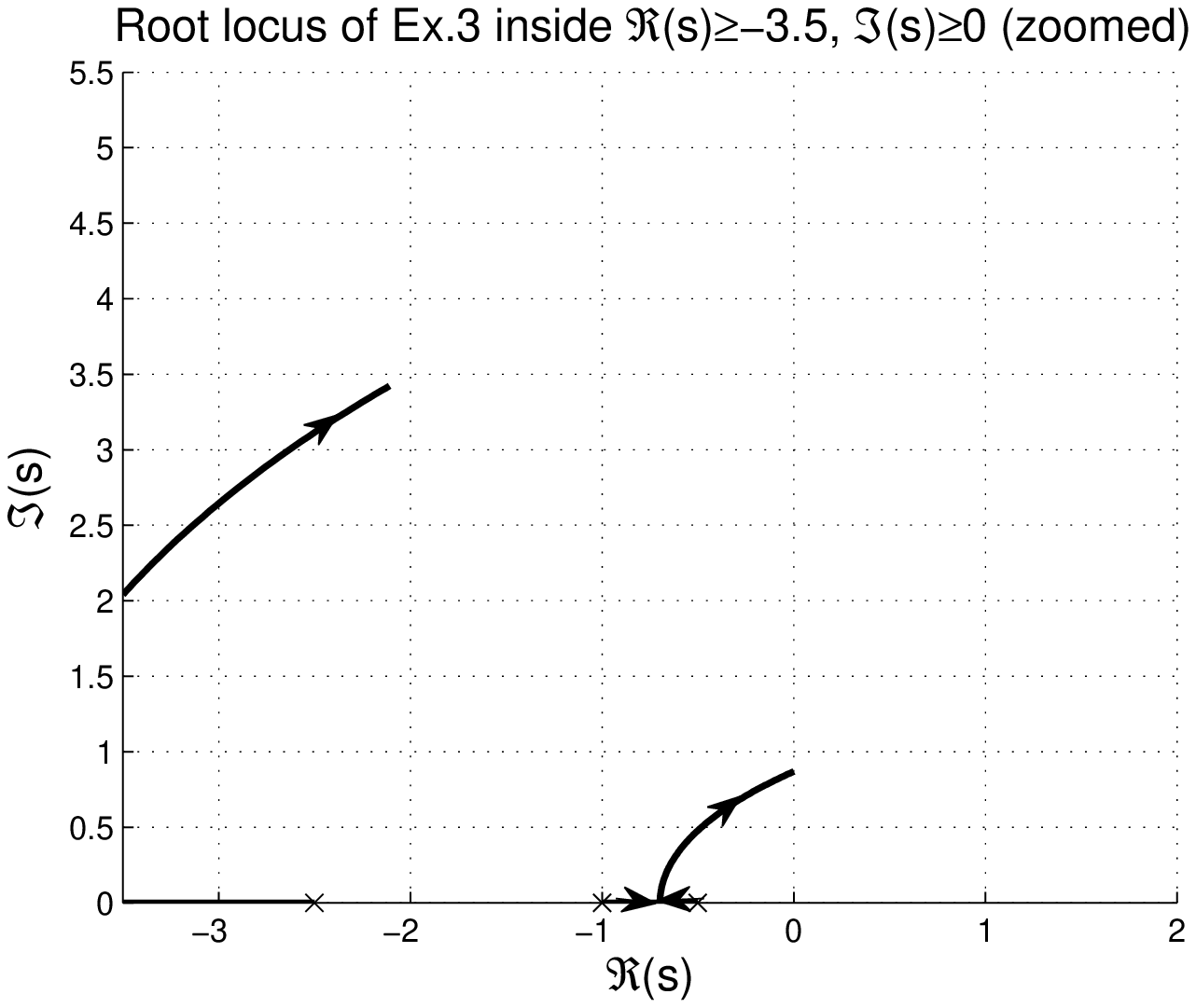}
\caption{\label{fig:rlocus3a} The root locus inside the gray region in Figure~\ref{fig:rlocus3} for controller gain $\lambda\in[0,0.07]$.}
\end{center}
\end{figure}

\begin{figure}[!h]
\begin{center}
\includegraphics[width=0.45\textwidth]{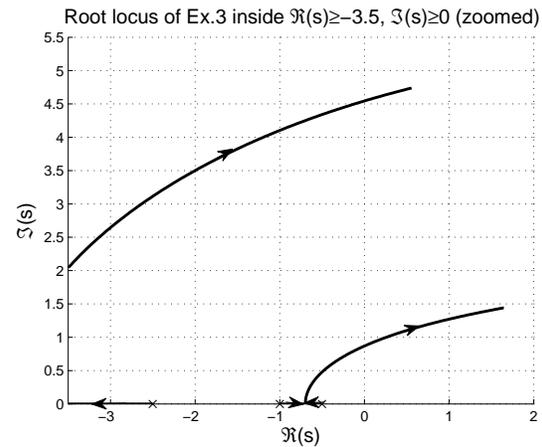}
\caption{\label{fig:rlocus3b} The root locus inside the gray region in Figure~\ref{fig:rlocus3} for controller gain $\lambda\in[0,5]$.}
\end{center}
\end{figure}

\begin{exmp} \label{ex3}
Finally, we consider the SISO dead-time system where $h=1$ and $G(s)=\frac{s^2-10s+50}{s^3 + 4 s^2 + 4.25 s + 1.25}$. We plot its root-locus trajectories of equation (\ref{eq:rlocusK}) for the controller gain interval $\lambda\in[0,5]$ inside $\Re(s)\geq-3.5$. Figure~\ref{fig:rlocus3} illustrates the general behavior of the root-locus trajectories. As the controller gain increases, more characteristic roots cross the boundary $\Re(s)=-3.5$ and asymptotic root trajectories get closer to the imaginary axis. Figure~\ref{fig:rlocus3a}-\ref{fig:rlocus3b} show the local behavior around the starting points, $s=-0.5, -1, -2.5$. In Figure~\ref{fig:rlocus3a}, we see that the characteristic roots cross the imaginary axis at $\lambda=0.07$ and the closed-loop system becomes unstable. In Figure~\ref{fig:rlocus3b}, the root locus has a branch point $s=-0.6976$ at $\lambda=0.0009$ and the trajectory starting with $s=-2.5$ leaves the region at $\lambda=0.0023$.
\end{exmp}
The computation times of root-locus plots for Examples \ref{ex1}-\ref{ex3} are $7.6$, $1.4$ and $2.5$ seconds on a PC with an Intel Core Duo $2.53$ GHz processor with $2$ GB RAM respectively.

\section{Concluding Remarks} \label{sec:concl}
The root locus of SISO dead-time systems with respect to the controller gain or the time delay is computed inside a given complex right half-plane. The method calculates the starting points of root-locus trajectories including the ones crossing the boundary of the root-locus region. Each trajectory is followed by a predictor-corrector continuation method. The characteristic roots on each root-locus trajectory are predicted by a secant method where the step length is adaptive and the predicted values are corrected by Newton's method.
 This continuation approach, which stems from numerical bifurcation analysis, is general and not restricted to parameterized equations  of the form in equations~(\ref{eq:rlocusK},\ref{eq:rlocusH}). The implementation of the method is numerically stable for high order systems.

\vspace{-0.35cm}

\begin{ack}
This work has been supported by the Programme of Interuniversity Attraction Poles of the Belgian Federal Science Policy Office (IAP P6- DYSCO), by OPTEC, the Optimization in Engineering Center of the K.U.Leuven, by the project STRT1-09/33 of the K.U.Leuven Research Council and the project G.0712.11N of the Research Foundation - Flanders (FWO).
\end{ack}

\vspace{-0.35cm}

\bibliographystyle{plain}
\bibliography{rlocus}
\end{document}